\documentclass[12pt, letterpaper, oneside]{article}

\usepackage{graphicx}
\usepackage[body={6.6in, 9in}, right=1in, top=1in]{geometry}
\usepackage{amssymb}
\usepackage{amsmath} 
\usepackage{feynmp} 
\usepackage{subfigure}
\usepackage{array}
\usepackage{bm}
\usepackage{cite}
\usepackage[linktoc=page,colorlinks=true,linkcolor=blue,citecolor=blue,urlcolor=blue]{hyperref}
\renewcommand{\l}{\lambda}
\newcommand{\m}{\mu}
\newcommand{\n}{\nu}

\newcolumntype{M}[1]{>{\centering\arraybackslash}m{#1}}

\def\be{\begin{equation}}
\def\ee{\end{equation}}
\def\ba{\begin{eqnarray}}
\def\ea{\end{eqnarray}}	
\def\h{\eta}

\def\del{\partial}

\newcommand{\sfrac}[2]{{\textstyle\frac{#1}{#2}}}

\begin{document}

\begin{fmffile}{graphs}

% TITLE AND AUTHOR

\begin{center}
\Large{\textbf{A technical analog of the cosmological \\
constant problem and a solution thereof}} \\[1cm]
\large{Ioanna Kourkoulou, Alberto Nicolis, and Guanhao Sun}
\\[0.4cm]

\vspace{.2cm}
\small{\em Center for Theoretical Physics and Department of Physics, \\
  Columbia University, New York, NY 10027, USA}

\end{center}

\vspace{.2cm}

% ABSTRACT

\begin{abstract}
The near vanishing of the cosmological constant is one of the most puzzling open problems in theoretical physics. We consider a system, the so-called framid, that features a technically similar problem. Its stress-energy tensor has a Lorentz-invariant expectation value on the ground state, yet there are no standard, symmetry-based selection rules enforcing this, since the ground state spontaneously breaks  boosts. We verify the Lorentz invariance of the expectation value in question with explicit one-loop computations. These, however, yield the expected result only thanks to highly nontrivial cancellations, which are quite mysterious from the low-energy effective theory viewpoint. 
\end{abstract}

%\newpage

% TABLE OF CONTENTS

%\tableofcontents

%\newpage

% INTRODUCTION

%%%%%%%%%%%%%%%%%%%%%%%%%%%%%%%%%%%%%%%%%%%
%%%%%%%%%%%%%%%%%%%%%%%%%%%%%%%%%%%%%%%%%%%
\tableofcontents

\section{Introduction}

The cosmological constant problem has been a topic of heated debate for decades. In fact, it is difficult to find two theoretical physicists who agree on what exactly the problem is, how to phrase it, how to quantify it, or how many problems there are. Some maintain that there is no problem at all. (We realize that some of our colleagues will take issue with this paragraph as well.)

We do not intend to enter the debate ourselves, nor to review the extensive literature on the subject
\footnote{It is impossible for us to do justice to all the attempts that have been made at tackling the cosmological constant problem. For an overview, we refer the reader to \cite{WeinbergCC, Bousso, Martin, Padilla, Hebecker} and references therein. In particular, ref.~\cite{Hebecker} reviews recent attempts based on relaxation mechanisms.}. Our aim with this paper is, instead, to describe a system that exhibits a technically similar problem, and to show in detail how that problem is ``solved" there. 
The quotes are in order: from a low-energy effective field theory viewpoint, the problem is solved by apparently miraculous cancellations. However, there is a more fundamental viewpoint, according to which those cancellations have to happen. Needless to say, it is  a symmetry that ultimately enforces those cancellations, but a symmetry that is spontaneously broken. In particular, that symmetry does not enforce the cancellations through standard selection rules---the ground state is not invariant under the symmetry---but only in a roundabout way, which is utterly obscure in the low-energy effective theory computation we will perform.

The system we have in mind is the so-called framid \cite{framid}. It is a hypothetical state of relativistic matter that spontaneously breaks Lorentz boosts but no other symmetry. The low-energy effective field theory involves three gapless Goldstone bosons $\vec \eta(x)$ that, like the broken boost generators they are associated with, transform as a vector under rotations. Their low-energy effective action can be written down systematically, in a derivative expansion, using for instance the coset construction for spacetime symmetries. From this, one can derive the stress energy tensor of the theory, which displays a very peculiar property: if evaluated on the $\vec \eta(x) = 0 $ background, it is Lorentz invariant \cite{framid}:
\be
T_{\mu\nu}(x)  = -\Lambda \, \eta_{\mu \nu} + {\cal O}(\partial \vec \eta \,) \; . \label{LI Tmn}
\ee
This is surprising: the ground state of the system spontaneously breaks boosts, so there is no obvious reason why it should have a Lorentz-invariant stress tensor. Certainly, all familiar condensed matter systems in the lab, such as solids, fluids, superfluids, which also spontaneously break Lorentz, have stress-energy tensors in their ground (or equilibrium) states that are {\em not} Lorentz invariant. In particular, in $c=1$ units, they typically have mass densities much bigger than their pressures or internal stresses.

So, what enforces the structure in \eqref{LI Tmn} for the framid? Unclear. It can be argued that it is the underlying Lorentz invariance of the theory, but only in the sense that if one writes down the most general framid effective action compatible with spontaneously broken Lorentz boosts, and then derives the corresponding stress-energy tensor, one ends up with eq.~\eqref{LI Tmn}. A more direct argument based on symmetry considerations directly for the expectation value of $T_{\mu\nu}$ is not available at the moment. In fact, there exists a completely different system that features precisely the same symmetry breaking pattern as the framid, but saturates the associated Goldstone theorem in a radically different way, and that most definitely does not have a Lorentz-invariant stress-energy tensor in its ground state. Such a system is the familiar Fermi liquid, and we refer the reader to \cite{AN} for details about its relationship with spontaneously broken boosts. So, apparently, symmetries cannot be the end of the story.

It could be that the structure in \eqref{LI Tmn} is only a tree-level statement, and gets modified upon taking into account quantum corrections. Concretely, this would mean that the expectation value of the stress-energy tensor on the framid's ground state is not Lorentz-invariant,
\be
\langle T_{\mu\nu}(x) \rangle \neq -\Lambda \, \eta_{\mu\nu} \, ,
\ee
for any $\Lambda$. This, however, does not seem to be consistent with renormalization theory and, more in general, renormalization group ideas. The reason is that we can think of our framid effective theory as a theory for physics below a certain energy scale $M$, with all the physics above $M$ having been integrated out. If we integrate out some more physics, say down to a scale $M' < M$, the coefficients in the framid low-energy effective action change, but the set of allowed terms in such an action remains the same. So, what was a ``quantum correction" in the energy window between $M'$ and $M$, now becomes a new contribution to a tree-level coupling in the low-energy theory. But at tree-level the effective theory yields
\be
\langle T_{\mu\nu}(x) \rangle = -\Lambda \, \eta_{\mu\nu} \; . \label{vev}
\ee
This suggests that the same should hold at the quantum level, up to a renormalization of $\Lambda$.

This is our technical analog of the cosmological constant problem. There, one has that the expectation value of the real world's stress-energy tensor that couples to gravity is zero (or fantastically smaller than the ``natural" value it should have), without any manifest symmetry reasons for why it should be so. In our case, we have that the Lorentz-violating components of the expectation value of the framid's stress-energy tensor are exactly zero, without any manifest symmetry reasons for why it should be so. 

Our paper is devoted to explicitly verifying eq.~\eqref{vev} for a framid to one-loop order. As one might expect, the computation will involve UV-divergent loop integrals, and we will have to pay particular attention to how we regulate such integrals. The reason is that our question has to do with Lorentz invariance, and so we need to make sure that our regulator respects it. However, since Lorentz invariance is spontaneously broken, it is not manifest in the effective theory or, more importantly, in our loop integrals. And so, for instance, cutting off our integrals in (Euclidean) momentum space in a manifestly Lorentz invariant fashion makes no sense: the framids' Goldstones already have propagation speeds different from unity, and from one another, and so why should their loops be cutoff in a Lorentz-invariant fashion?

To address this problem, we use two regulators that can be made straightforwardly compatible with our spontaneously broken Lorentz invariance. The first is a generalization of Pauli-Villars. The second is dimensional regularization. Notice that, for the latter, since the Goldstone's effective theory has no mass parameters, the relevant integrals for our one-loop computation will all be trivially zero, which would make our check moot. To circumvent this, we couple the framid to a massive scalar particle, in all ways allowed by symmetry, and run our check in that case.

With both regulators, we find that, indeed, eq.~\eqref{vev} is obeyed at one-loop order. However, the cancellations involved in making the Lorentz-violating components vanish are absolutely nontrivial, and we were not able to find a clear structure in the actual computation that would ensure, or even suggest, those cancellations. As a further check of our techniques and of the non-triviality of the cancellations, we 
run the same computation in the case of a superfluid, where we {\em don't} expect analogous cancellations---a superfluid certainly does not respect \eqref{vev}. Indeed, in that case we find deviations from \eqref{vev}, and our result matches precisely that of an independent computation carried out elsewhere \cite{JNPS}.

\vspace{.5cm}
\noindent
{\em Notation and conventions:} We use natural units ($\hbar=c=1$) and the mostly-plus metric signature throughout the paper.

%%%%%%%%%%%%%%%%%%%%%%%%%%%%%%%%%%
%%%%%%%%%%%%%%%%%%%%%%%%%%%%%%%%%%
\section{Warm-up: the  free relativistic scalar}\label{section2}

Before attempting to compute the expectation value of $T^{\mu\nu}$ for our framid, it is instructive to review how things work out for a free relativistic scalar field, even if we perform computations in a way that is {\em not} manifestly Lorentz covariant. 

Consider thus the Lagrangian
\be
{\cal L} =  -\sfrac12(\partial \phi)^2 -  \sfrac12 m^2 \phi^2  \; , \label{relativistic scalar}
\ee
which yields the stress-energy tensor
\be
T^{\mu\nu} = \partial^\mu \phi \partial^\nu \phi + \eta^{\mu\nu} {\cal L} \; .
\ee

A quadratic Lagrangian can always be rewritten as a total derivative plus a term proportional to the equations of motion.
So, since the vacuum is translationally invariant, the second term in $T^{\mu\nu}$ does not contribute to our expectation value, and we simply have
\begin{align}
\langle T^{\mu\nu} (x) \rangle & = \langle \partial^\mu \phi(x) \partial^\nu \phi(x) \rangle \label{Tmn vev}\\
& = -\lim_{y \to x} \partial_x^\mu \partial_x^\nu G_W(x-y) \\
& = \int \frac{d^4 k}{(2 \pi)^4} \, {k^\mu k^\nu} \tilde G_W(k)\; ,  \label{k k GW}
\end{align}
where $G_W$ is the Wightman two-point function of $\phi$:
\be
\tilde G_W(k) = \theta(k^0) \, (2 \pi) \delta(k^2+m^2) \; .
\ee 

At this stage one could use Lorentz-invariance and conclude that
\be
\langle T^{\mu\nu} (x) \rangle = \frac14 \eta^{\mu\nu} \int \frac{d^4 k}{(2 \pi)^4} \, k^2 \tilde G_W(k) \; .
\ee
The integral is UV divergent and must be regulated. However, whatever its value, we see that $\langle T^{\mu\nu} (x) \rangle$ is proportional to $\eta^{\mu\nu}$, as expected.

Let's instead go back to eq.~\eqref{k k GW}, give up manifest Lorentz invariance, and use the delta function in $\tilde G_W$ to perform the integral over $k^0$. Using spatial rotational invariance, we get
\be
\rho \equiv \langle T^{00} \rangle = \frac12 \int \frac{d^3 \textbf{{k}}}{(2 \pi)^3} \, \omega_k \; , 
\qquad \langle T^{0i}  \rangle = 0 \;, 
\qquad p \equiv \frac13 \langle T^{ii} \rangle = \frac16 \int \frac{d^3 \textbf{{k}}}{(2 \pi)^3} \frac{\textbf{{k}} ^2}{\omega_k} \; , \label{rho p relativistic}
\ee
where $\omega_k = \sqrt{\textbf{{k}}  ^2+m^2}$. $\langle T^{\mu\nu} \rangle$ is proportional to $\eta^{\mu\nu}$ if and only if $\rho + p$ vanishes, but now this seems impossible: the integrals entering $\rho$ and $p$ are manifestly positive definite, so, how can there be any cancellations? Then again, such integrals are UV divergent, so one should regulate them properly before jumping to conclusions.

The UV regulator used should preserve Lorentz invariance. In particular, a hard cutoff in momentum space will not do, because we have already performed the integral in $k^0$, and so at this point there is no way to introduce a hard cutoff compatible with Lorentz invariance (usually this involves Wick-rotating the $d^4k$ integral to Euclidean space, and then imposing a 4D rotationally invariant cutoff there.) 

One possibility is to use dimensional regularization directly for our $3$-dimensional integrals. This is compatible with Lorentz invariance because it corresponds to formulating the original theory \eqref{relativistic scalar} in $d+1$-dimensions, going through the manipulations \eqref{Tmn vev}-\eqref{k k GW} in $d+1$ dimensions, and then performing the $k^0$ integral explicitly to be left with $d$-dimensional integrals. We get:\footnote{Here and for the rest of the paper we avoid introducing the MS renormalization scale $\mu$, which would be needed to make our dim-reg formulae dimensionally correct. We do so for notational simplicity, to avoid clutter. If one wants to reinstate $\mu$ in our formulae, one can do so just by dimensional analysis, interpreting our formulae as being expressed in $\mu=1$ units.}
\begin{align}
\rho & = \frac12 \int \frac{d^d \textbf{{k}}}{(2 \pi)^d} \, \omega_k = -m^{d+1} \frac{\Gamma\big(-\frac{d+1}{2}\big)}{2(4\pi)^{\frac{d+1}{2}}} \label{rho dim reg} \\
p & =  \frac1{2d} \int \frac{d^d \textbf{{k}}}{(2 \pi)^d} \frac{\textbf{{k}} ^2}{\omega_k} = m^{d+1} \frac{\Gamma\big(-\frac{d+1}{2}\big)}{2(4\pi)^{\frac{d+1}{2}}} \label{p dim reg} 
\end{align}
(notice the $\frac13 \to \frac 1 d$ replacement in the definition of $p$), in agreement with $\rho+p =0$.

Another possibility is to use a generalization of Pauli-Villars. Recall that, in the simplest case, Pauli-Villars amounts to regulating a log-divergent loop integral by a modification of the Feynman propagator of the form
\be
\tilde G_F(k) = \frac{i}{-k^2-m^2+i\epsilon} \to \tilde G^{PV}_F(k) = \frac{i}{-k^2-m^2+i\epsilon} - \frac{i}{-k^2-M^2+i\epsilon} \; ,
\ee
with $M$ being a very large mass scale, in particular $M \gg m$. This improves the UV behavior of the integral without affecting the IR one:
\be \label{GPV}
\tilde G^{PV}_F(k \ll M) \simeq \tilde G_F(k) \; , \qquad \tilde G^{PV}_F(k \gg M) \simeq -\frac{i(M^2 - m^2)}{k^4} \; .
\ee
This is manifestly a Lorentz invariant modification of the propagator. More importantly for our applying these ideas to the framid case, such a modification corresponds to adding certain Lorentz-invariant higher derivative terms to the Lagrangian, as clear from the exact rewriting (up to $i \epsilon$'s)
\be \label{GPV combined}
\tilde G^{PV}_F(k) = \frac{i}{-\frac{M^2+m^2}{M^2-m^2} \cdot k^2 - \frac{1}{M^2-m^2} \cdot k^4 - \frac{M^2}{M^2-m^2} \cdot m^2} \; ,
\ee
or, keeping only the leading order in $m/M$, %should this be m/M? (GS)
\be
\tilde G^{PV}_F(k) \simeq \frac{i}{- k^2 - k^4/M^2 -  m^2} \; .
\ee

The introduction of the Pauli-Villars propagator is enough to regulate log-divergent integrals. In our case, we have quartically divergent integrals, with subleading quadratic and log divergences as well. It turns out that to cancel all these divergences, we need a three-fold modifications of the Feynman propagator. That is, calling $\tilde G_{F}(k; M)$ the Feynman propagator for generic mass $M$, we want to perform the replacement
\be
\tilde G_{F} (k; m) \to \tilde G^{PV}_F(k) = \tilde G_{F} (k; m) + \sum_{a=1}^3 c_a \, \tilde G_{F} (k; \alpha_a M) \; ,
\ee
where the $c$'s and $\alpha$'s are suitable (order-one) coefficients, and  $M$ is a common  very large mass scale, $M \gg m$. 

Since we need to improve the high energy behavior of our loop integrals by four powers of $k$ compared to the standard Pauli-Villars case of \eqref{GPV}, we want our modified propagator to have the high energy behavior
\be
\tilde G^{PV}_F(k \gg M) \sim \frac{1}{k^8} \; ,
\ee
which requires the $\alpha_a$'s to be all different, and the $c_a$'s to be
\be
c_a =  \prod_{b\neq a}\frac{\alpha_b^2 -\frac{m^2}{M^2}}{\alpha^2_b - \alpha^2_a}  \; . \label{c_i}
\ee
By combining denominators as done above in \eqref{GPV combined}, one can see that this still corresponds to adding suitable higher-derivative Lorentz-invariant terms to the Lagrangian.

Our expressions for $\rho$ and $p$ in eq.~\eqref{rho p relativistic} do not involve directly a Feynman propagator. However, we have to remember that they came from integrating over $k^0$ an expression involving the Wightman two-point function of $\phi$. In our case, we must replace this with
\be
\tilde G_{W} (k; m) \to \tilde G^{PV}_W(k) = \tilde G_{W} (k; m) + \sum_{a=1}^3 c_a \, \tilde G_{W} (k; \alpha_a M) .
\ee
Then, integrating again over $k^0$, for $\rho$ and $p$ we simply get
\begin{align}
\rho & = \frac12 \int \frac{d^3 \textbf{{k}}}{(2 \pi)^3} \, \big[\omega_{k; \, m} + \sum_{a=1}^3 c_a \, \omega_{k; \, \alpha_a M} \big] \\
p & = \frac16 \int \frac{d^3 \textbf{{k}}}{(2 \pi)^3} \, \textbf{{k}} ^2 \bigg[ \, \frac1{\omega_{k; \, m}} 
+\sum_{a=1}^3 c_a \frac1{ \omega_{k; \, \alpha_a M}} \, \bigg] \; , \qquad  \omega_{k; \, M} \equiv \sqrt{\textbf{{k}}^2 + M^2} \; .
\end{align}
As expected, but still surprisingly enough, choosing the $c_a$'s as in eq.~\eqref{c_i} makes both of these integrals finite, and, in fact, opposite to each other. Namely:
\be
\rho = - p = f(\alpha) M^4 + g(\alpha) m^2 M^2  + \frac{1}{32\pi^2} m^4 \log(m/M) + h(\alpha) m^4 \; ,
\ee
where $f$, $g$, and $h$ are somewhat complicated functions of the $\alpha$ coefficients, whose explicit form we spare the reader. On the other hand,  the coefficient of $m^4 \log m$ is finite (for $M\to\infty$) and $\alpha$-independent, in agreement with standard renormalization theory: like all non-analyticities in external momenta and mass parameters, it should be finite and calculable, that is, independent of the regulator used. In fact, if we expand the dim-reg result \eqref{rho dim reg} for $d \to 3$, we get exactly the same coefficient for $m^4 \log m$.

%%%%%%%%%%%%%%%%%%%%%%%%%%%%%%%%%%%%%%%%%%%
%%%%%%%%%%%%%%%%%%%%%%%%%%%%%%%%%%%%%%%%%%%
\section{The framid stress-energy tensor}\label{framid}
The low-energy effective theory for the framid, a relativistic system that spontaneously breaks Lorentz boosts, was first developed in \cite{framid}. Here, we provide a brief description of the theory to set the ground for our computation. 

Framids are most intuitively described in terms of  a  vector field $A_\mu(x)$ with a constant time-like expectation value:
\begin{equation}
 \langle A_\mu(x)\rangle=\delta_\mu^0 \; .
\end{equation}
Such an expectation value breaks Lorentz boosts, and the corresponding Goldstone fields $\vec\eta(x)$ can be thought of as parametrizing the fluctuations of $A_\mu(x)$ in the directions of the broken symmetries:
\begin{equation}
A_\mu(x) = \big(e^{i\vec\eta(x)\cdot \vec K }\big)_\mu {}^{\alpha}\langle A_\alpha(x)\rangle \; , \label{3}
\end{equation}
where $\vec K$ are the boost generators. Notice that, even with Goldstone fields present, $A_\mu A^\mu = -1$.

One can construct the Goldstone effective theory either by writing down in a derivative expansion the most general Poincar\'e invariant action for $A_\mu$, performing the replacement \eqref{3}, and then expanding to any desired order in the $\vec \eta$ fields, or by using the coset construction for spontaneously broken spacetime symmetries. These two approaches have different advantages and disadvantages, but they yield the same result \cite{framid}. We will follow the former.

To the second order in the derivative expansion, the most general effective Lagrangian takes the form \cite{framid}
\begin{equation}
    \mathcal{L} = -\frac{ M_1^2}{2}\left[(c_L^2-c_T^2)(\partial_\mu A^\mu)^2 + c_T^2(\partial_\mu A_\nu)^2 + (c_T^2-1)(A^\rho\partial_\rho A_\mu)^2\right] \; , \label{2}
\end{equation}
where $M_1$ is an overall mass scale and $c_L, c_T$ represent the propagation speeds of the longitudinal and transverse Goldstones. 
We want to eventually work with the Goldstone Lagrangian expanded to quadratic order, hence, with two derivatives in each term, it suffices to expand $A_\mu$ to first order in the Goldstones,
\begin{equation}
A_0 = \Lambda_0^{\;0} = \cosh|\vec\eta \, | \simeq 1,\quad\quad A_i = \Lambda_i^{\;0}=\frac{\eta^i}{|\vec\eta \, |}\sinh{|\vec\eta \,|} \simeq\eta^i \; .
\end{equation}

It is convenient to separate the Goldstones into their longitudinal and transverse modes, $\vec\eta=\vec\eta_L+\vec\eta_T$, where
\begin{equation}
\vec\nabla\times\eta_L=0,\quad\quad \vec\nabla\cdot\vec\eta_T=0 \; ,
\end{equation}
and also rescale them,
\begin{equation}
\vec\eta \to \frac{\vec\eta}{M_1},
\end{equation}
  to eventually obtain a neat form of the effective Goldstone Lagrangian,
\begin{equation} \label{L2}
    \mathcal{L}_2= \frac{1}{2}\left[\dot{\vec\eta} \, ^2-c_L^2 \, (\vec\nabla\cdot\vec\eta_L)^2-c_T^2 \, (\partial_i\eta_T^j )^2 \right].
\end{equation}

Our goal is to check whether the stress-energy tensor resulting from this Lorentz-violating theory remains Lorentz-invariant when including quantum corrections: $\langle T_{\mu\nu}(x)\rangle = -\Lambda \,\eta_{\mu\nu}$. We know the off-diagonal components of the stress-energy tensor respect this condition due to rotational invariance of the ground state, hence our task reduces to proving that 
\begin{equation}
    \langle T^{00}(x)\rangle +  \frac{1}{3}\langle T^{ii}(x)\rangle=0 \; ,
\end{equation} 
where the spatial indices are implicitly summed over (as for the rest of the paper).

In order to compute the one-loop correction to $\langle T^{\mu\nu}(x)\rangle$, we start from the full covariant theory \eqref{2} and compute the canonical (Noether) stress-energy tensor,
\begin{align}
T^{\m\n} &=  -\frac{\del\mathcal{L}}{\del(\del_{\m}A_{\l})}\del^{\n}A_{\l} + g^{\m\n}\mathcal{L} \nonumber\\
&= M_1^2\left[ (c_L^2-c_T^2)\del^\n A^\m\del_\l A^\l + c_T^2\del^\m A_\l\del^\n A^\l + (c_T^2-1)A^\m\del^\n A_\l A^\rho\del_\rho A^\l\right] + g^{\m\n}\mathcal{L} . \label{8}
\end{align}
 
We wish to introduce the framid Goldstones as in \eqref{3} and expand $T^{\m\n}$ up to quadratic order. The diagonal components of \eqref{8} reduce to
\begin{align}
T^{00}  
&= -\mathcal{L}+\dot{\vec\h}\cdot\dot{\vec\h}, \\
T^{ii}&=3\mathcal{L} + c_T^2\del_i\vec\h\cdot\del_i\vec\h +(c_L^2-c_T^2)(\vec\nabla\cdot\vec\h)^2 \; .
\end{align}
Not surprisingly, this is the same result we would have gotten by applying Noether's theorem directly to eq.~\eqref{L2}: even though boosts are spontaneously broken, spacetime translations are not, and so one can compute the associated Noether current using directly the $\vec \eta$ parametrization of the action and having $\vec \eta$ transform in the usual way under translations, $\vec \eta \to \vec \eta  - \epsilon^\mu \partial_\mu \vec \eta$.

We now perform manipulations similar to those of sect.~\ref{section2}. Dropping the terms proportional to the Lagrangian for the same reason as made explicit there, the ground-state expectation values of the expressions above can be written as
\begin{align}
\rho & \equiv \langle T^{00} \rangle = \lim_{y \rightarrow x} 
\partial_{t_x} \partial_{t_y} \langle \vec{\eta}(x)\cdot\vec{\eta}(y) \rangle  \label{rho framid} \\ 
p & \equiv \frac{1}{3} \langle T^{ii} \rangle = \frac{1}{3} \lim_{y \rightarrow x} \Big[ c_T^2 \, \partial^i_x  \partial^i_y \langle \vec{\eta}(x)\cdot\vec{\eta}(y) \rangle + (c_L^2 - c_T^2) \partial^i_x  \partial^j_y \langle {\eta}^i(x) \eta^j(y) \rangle \Big] \label{p framid} 
\end{align}
Decomposing $\vec{\eta}$ into longitudinal and transverse components,  the energy density becomes
\begin{align}
\rho =&  - \lim_{y \rightarrow x}\partial_{t_x}^2  \langle \vec{\eta}_L(x)\cdot\vec{\eta}_L(y) + \vec{\eta}_T(x)\cdot\vec{\eta}_T(y) \rangle \nonumber\\
=& -\lim_{y \rightarrow x}\partial_{t_x}^2 \left( G_L(x-y) + 2 \, G_T(x-y) \right) \nonumber\\
=& \frac12 \int \frac{d^3 \textbf{{k}} }{(2\pi)^3} \left( \omega_L + 2 \, \omega_T \right), \label{38}
\end{align}
where $G_L$ and $G_T$ are the (scalar) Wightman two-point functions for longitudinal and transverse modes, 
\be \label{GLT}
\tilde G_{L/T}(\omega, \textbf{{k}} \,) = \theta(\omega) (2 \pi) \delta(\omega^2 - \omega_{L/T}^2) \; ,
\ee
and $\omega_L$ and $\omega_T$ are the corresponding energies,  $\omega_{L/T} = c_{L/T}\, | \textbf{{k}} |$. The relative factor of two in \eqref{38} comes from the fact that $\hat{k}\cdot\hat{k} = 1$ and $\sum_i \delta^{ii} - \hat{k}^i\hat{k}^i = 2$. %Use k instead of p? (GS)

Similarly, the pressure can be rewritten as 
\be
p = \frac{1}{6}  \int \frac{d^3 \textbf{{k}} }{(2\pi)^3} \left( \frac{c_L^2 \, \textbf{{k}}^2}{\omega_L} +  2 \frac{c_T^2 \; \textbf{{k}}^2}{\omega_T} \right).\label{39}
\ee

These expressions for $\rho$ and $p$ are very simple generalizations of the corresponding ones in sect.~\ref{section2} for a generic massive scalar. However, as we did there, we first need to regularize them before we can check whether $\rho+p$ vanishes. As we emphasized in the Introduction, the regulators used should be consistent with the spontaneously broken Lorentz invariance.

%%%%%%%%%%%%%%%%%%%%%%%%%%%%%%%%%%%%%%%%%%%
%%%%%%%%%%%%%%%%%%%%%%%%%%%%%%%%%%%%%%%%%%%
\section{Pauli-Villars regularization}

Let us first consider a suitable generalization of Pauli-Villars regularization. Given the technical similarities between the derivation we just performed and that of sect.~\ref{section2}, it's clear that {\em if} we are allowed to introduce independent Pauli-Villars modifications for the longitudinal and transverse phonons' propagators, taking into account their different speeds, then we get $\rho+p = 0$ for the framid as well.  

More explicitly, consider the longitudinal phonons' contributions to $\rho$ and $p$,
\be
\rho_L \equiv \frac12 \int \frac{d^3 \textbf{{k}} }{(2\pi)^3} \omega_L \; ,   \qquad p_L \equiv   \frac{1}{6}  \int \frac{d^3 \textbf{{k}} }{(2\pi)^3}  \frac{c_L^2 \, \textbf{{k}}^2}{\omega_L} \; ,
\ee 
with $\omega_L = c_L |\textbf{{k}}|$. Apart from the integration measure, $\textbf{{k}}$ always appears here in the combination $c_L |\textbf{{k}}|$. The same is true for the Wightman two-point function \eqref{GLT}, where these expressions come from, and for the associated Feynman propagator,
\be
\tilde G^L_F(\omega, \textbf{{k}}) = \frac{i}{\omega^2 - c_L^2 \textbf{{k}} \,^2 + i \epsilon } \; .
\ee
So, upon changing the integration variable, $\textbf{{k}} \, ' = c_L \textbf{{k}}$, we have
\be
\rho_L = \frac{1}{c^3_L} \rho_{\rm rel}(0) \; , \qquad p_L = \frac{1}{c^3_L} p_{\rm rel}(0) \; ,
\ee
where $\rho_{\rm rel}(m)$ and $p_{\rm rel}(m)$ are the  relativistic expressions for the energy density and pressure of a massive scalar of mass $m$, eq.~\eqref{rho p relativistic}. Then, the same Pauli-Villars regularization that was applied in sect.~\ref{section2} can be applied here, yielding
\be
\rho_L+ p_L = 0 \; .
\ee
 Mutatis mutandis, the same considerations can be applied to the transverse phonons' contributions $\rho_T$ and $p_T$, yielding
\be
\rho_T+ p_T = 0 \; .
\ee
And so, combining the longitudinal and transverse sectors, $\rho+p = 0$.

Notice however that the longitudinal and transverse Goldstones have different speeds in general, and so the Pauli-Villars regularization procedure that we are advocating has to be different for the two sectors. Namely, referring to the explicit analysis of sect.~\ref{section2}, whenever there is a $k^2$ in a propagator, when we deal with the longitudinal sector we have to replace that with $-\omega^2 + c_L^2 \textbf{{k}} \, ^2$, and when we deal with the transverse sector we have to replace that with $-\omega^2 + c_T^2 \textbf{{k}} \, ^2$.

Recall that in standard relativistic cases, such as that studied in sect.~\ref{section2}, a Pauli-Villars modification of the propagator can be thought of as coming directly from a suitable local higher-derivative modification of the action.
So, in our case the question is whether there is a local and Lorentz-invariant higher-derivative modification of the framid's action that corresponds to {\em independent} Pauli-Villars modifications of the longitudinal and transverse propagators of the desired type. The requirements of locality and Lorentz-invariance are nontrivial---the former because the longitudinal/transverse splitting of $\vec \eta$ is non-local, the latter because $\vec \eta$ transforms nonlinearly under Lorentz boosts.

Let's start with locality. The modifications of the Feynman propagators we are after, upon combining denominators as explained in sect.~\ref{section2}, take the form
\be
\tilde G^{L, PV}_F(\omega, \textbf{{k}}) = \frac{i}{-k_L^2 - \frac{1}{\Lambda_1^2} k_L^4 - \frac{1}{\Lambda_2^4} k_L^6 - \frac{1}{\Lambda_3^6} k_L^8} \; , \qquad k^2_L \equiv -\omega^2 + c_L^2 \textbf{{k}} \,^2 \; ,
\ee
and similarly for the transverse propagator. The $\Lambda_a$'s are suitable combinations of the Pauli-Villars pole masses. Importantly for what follows, they are the same for the longitudinal and transverse propagators, as long as the pole masses are chosen to be the same for the two sectors. This can be understood easily by thinking about the structure of the propagators at $\textbf{{k}} = 0$, in which case the fact that the longitudinal and transverse propagation speeds are different does not matter.

Using the canonical normalization of \eqref{L2}, up to total derivatives these propagators correspond to the quadratic Lagrangian
\be \label{L2PV}
{\cal L}_2^{PV} = \frac12 \, \vec \eta_L \cdot \Big[ \Box_L - \frac{1}{\Lambda_1^2} \Box_L^2 + \frac{1}{\Lambda_2^4} \Box_L^3 - \frac{1}{\Lambda_3^6} \Box_L^4 \Big] \vec \eta_L + (L \to T) \; ,
\ee
where $\Box_L$ denotes the differential operator
\be
\Box_L \equiv -\partial_t^2 + c_L^2 \nabla^2 \; ,
\ee
and `$(L \to T)$' stands for a similar structure involving the transverse field $\vec \eta_T$ and its propagation speed. Our question of locality is thus reduced to the question of whether this quadratic Lagrangian can be written as a local quadratic Lagrangian for the full $\vec \eta$ field.  

Notice that as long as at least one spatial laplacian acts on $\eta_L$ or $\eta_T$, one can easily perform the longitudinal/transverse decomposition in a local fashion: defining the local differential operator matrix,
\be
D_{ij} = \partial_i \partial_j \; ,
\ee
we simply have
\be
\nabla^2 \vec \eta_L = D \cdot \vec \eta \; , \qquad \nabla^2 \vec \eta_T = (\nabla^2 - D)\cdot \vec \eta \; .
\ee 
Notice also that, in the quadratic action \eqref{L2PV}, it is enough that the $\vec \eta \,$'s on the right---those acted upon by derivatives---be split into longitudinal and transverse. The undifferentiated $\vec \eta \,$'s on the left can be replaced with the full $\vec \eta$ field, because, as usual, at quadratic order all longitudinal-transverse mixings automatically vanish.

So, the only remaining question concerning locality is whether the terms in \eqref{L2PV} with time-derivatives {\em only} can be rewritten in a local fashion. However, the coefficients of such terms are the same for the longitudinal and transverse sectors. This, upon using again the vanishing of  longitudinal-transverse mixings, allows us to combine these terms into purely time-derivative terms for the full $\vec \eta$ field:
\begin{align}
{\cal L}_2^{PV} & \supset \frac12 \, \vec \eta_L \cdot \Big[ -\partial_t^2 - \frac{1}{\Lambda_1^2} \partial_t^4 - \frac{1}{\Lambda_2^4} \partial_t^6 - \frac{1}{\Lambda_3^6} \partial_t^8 \Big] \vec \eta_L + (L \to T) \\
& = \frac12 \, \vec \eta \cdot \Big[ -\partial_t^2 - \frac{1}{\Lambda_1^2} \partial_t^4 - \frac{1}{\Lambda_2^4} \partial_t^6 - \frac{1}{\Lambda_3^6} \partial_t^8 \Big] \vec \eta
\end{align}

So, in summary, a local rewriting of \eqref{L2PV} involves the following building blocks:
\be \label{blocks}
\vec \eta \cdot \partial_t^{2 a} (\nabla^2)^b D^{c} \cdot \vec \eta \; ,
\ee
with different integer non-negative values for $a$, $b$, and $c$, up to $a+b+c=4$, and suitable coefficients. In particular, we can restrict to $c=0,1$, because $D$ is proportional to a projector operator (its only role is to isolate the longitudinal component of $\vec \eta \, $):
\be
D \cdot D =\nabla^2 \cdot  D  \; .
\ee
We can now ask whether these building blocks are compatible with the spontaneously broken Lorentz invariance. In particular, we can ask whether there exist manifestly Lorentz-invariant combinations of $A_\mu$ and $\partial_\mu$ that, when expanded to quadratic order in the $\vec \eta$ fields, reduce precisely to these building blocks.

It is quite easy to convince oneself that the answer is yes. To this end, it is convenient to integrate by parts half of the derivatives in \eqref{blocks}. Then, up to a possible sign, the building blocks are 
\be \label{c=0}
\big((\partial_t)^a \partial_{i_1} \dots \partial_{i_b} \eta^k \big)^2 \qquad \qquad(c=0)
\ee
and
\be \label{c=1}
\big( (\partial_t)^a \partial_{i_1} \dots \partial_{i_b} (\vec \nabla\cdot \vec \eta \, ) \big)^2 \qquad  (c=1)
\ee
Recalling that, to first order in $\vec \eta$, $A_\mu$ is
simply
\be
A_0 \simeq 1  \; , \qquad \vec  A \simeq \vec \eta \; ,
\ee
we see that two simple Lorentz-invariant generalizations of \eqref{c=0} and \eqref{c=1} that reduce to them to quadratic order in $\vec \eta$ are
\be \label{c=0 A}
\big((\partial^\parallel)^a \partial^\perp_{\mu_1} \dots \partial^\perp_{\mu_b} A_\nu \big)^2 \qquad \qquad(c=0)
\ee
and
\be \label{c=1 A}
\big( (\partial^\parallel)^a \partial^\perp_{\mu_1} \dots \partial^\perp_{\mu_b} (\partial_\nu A^\nu ) \big)^2  \; ,  \qquad  (c=1)
\ee
where $\partial^\parallel$ and $\partial^\perp_{\mu}$ are defined as
\be
\partial^\parallel \equiv -A^\alpha \partial_\alpha \; , \qquad \partial^\perp_{\mu} \equiv  \partial_{\mu} + A_\mu A^\alpha \partial_\alpha \; .
\ee

In conclusion: there exist  local Lorentz-invariant higher-derivative corrections to the framid action that modify the $\vec \eta$ propagator in a Pauli-Villars fashion, with suitable independent modifications for the longitudinal and transverse modes, such that the Pauli-Villars analysis of sect.~\ref{section2} can be separately applied to the two sectors, yielding $\rho + p = 0$.

\section{Dimensional regularization} 
Let's now consider dimensional regularization. As for the relativistic case considered in sect.~\ref{section2}, dimensional regularization of our spatial momentum integrals \eqref{38} and \eqref{39} is consistent with Lorentz invariance. This is because we can think of it as corresponding to formulating the original manifestly Lorentz-invariant theory for $A_\mu$, eq.~\eqref{2}, in $d+1$ spacetime dimensions, and then going through all the subsequent steps that led us to \eqref{38} and \eqref{39} keeping $d$ generic. Were we to  do so, we would end up with the same integrals \eqref{38} and \eqref{39}, but in $d$ rather than $3$ dimensions, and with a $1/2d$ rather than $1/6$ prefactor for the pressure.

Now, since our effective theory does not feature mass parameters for the Goldstone fields, the integrals \eqref{38} and \eqref{39} vanish in dimensional regularization. This is certainly consistent with $\rho+p = 0$, but in a trivial way. To run a nontrivial check, we need to deform our theory. % "deform our theory"? (GS)
We must do so consistently with Lorentz invariance, of course. And so, in particular, mass parameters for the Goldstones are not allowed. %"so in particular" -> "in particular"? (GS)

A particularly physical way to deform the theory is to couple the framid to a massive field, in all ways allowed by symmetry. For simplicity, let's take this to be a scalar field, $\phi$. If its mass is below the cutoff of the effective theory (the $M_1$ of sect.~\ref{framid}, up to suitable powers of $c_{L/T}$), $\phi$ must be included in our computation. Assuming it has zero expectation value, for our one-loop computation we are interested in all Lorentz-invariant combinations of $A_\mu$ and $\phi$ with up to two derivatives and which, once expanded about the framid's background configuration, yield quadratic terms in the $\vec \eta$ and $\phi$ fields.

% we couple our theory to a massive scalar field $\phi$ in all ways allowed by symmetry. Namely, we are interested in all possible terms at the same order of the derivative expansion (at most two derivatives) that yield quadratic terms in the Goldstone effective Lagrangian. 
 
We find that there are only three interactions with these properties:
\begin{equation}
    \mathcal{L} \quad\supset \quad A_\mu\partial^{\mu}\phi, \quad A_{\mu}A_{\nu}\partial^{\mu}\partial^{\nu}\phi, \quad A_{\mu}A_{\nu}\partial^{\mu}\phi\partial^{\nu}\phi \; .
\end{equation}
All other possibilities are either related to these through integration by parts or, when expanded in the Goldstone fields to the desired order, yield terms that are total derivatives themselves, and can thus be neglected. 
Adding the terms above to the framid action, and including also standard kinetic and mass terms for $\phi$, which are Lorentz-invariant by themselves, our effective Lagrangian at quadratic order becomes 
\begin{align}
    \mathcal{L}_2 \quad \to \quad  \frac{1}{2}\Big[ & \dot{\vec\eta} \, ^2-c_L^2 \, (\vec\nabla\cdot\vec\eta_L)^2-c_T^2 \, (\partial_i\eta_T^j )^2 +  \dot{\phi}^2 -(\vec\nabla \phi)^2 -m^2\phi^2  \nonumber\\
    & + 2 b_1 \, \phi \,\vec\nabla\cdot\vec{{\eta}}_L +
    2 b_2 \, \dot{\phi}\,\vec\nabla\cdot\vec{\eta}_L + b_3 \, \dot\phi^2 \Big] \label{52},
\end{align}
where the $b_a$'s are generic coupling constants. Notice that the transverse components of the Goldstone fields don't mix with the massive scalar, since $\vec\nabla\cdot\vec\eta_T=0$. 

The relevant components of the stress-energy tensor 
%can be computed by applying Noether's theorem to the action \eqref{52}, and take the following form: 
now read
\begin{align}
    T^{00}&=  \dot{\vec{\eta}} \, ^2 + b_2 \, \dot\phi \, \vec\nabla\cdot\vec{{\eta}}_L + (1 + b_3)\dot\phi^2, \label{53} \\
    T^{ii} &= c_L^2(\vec\nabla\cdot\vec\h_L)^2  + c_T^2 \, \del_i\vec\h_T\cdot\del_i\vec\h_T  - b_1 \phi \,\vec\nabla\cdot\vec{{\eta}}_L - b_2 \dot{\phi}\,\vec\nabla\cdot\vec{\eta}_L + (\vec\nabla \phi)^2 \label{54},
\end{align}
where we have omitted  terms proportional to the Lagrangian, since, as before, they do not contribute to our expectation values.

Notice that, at this order, the transverse Goldstones $\vec \eta_T$ are completely decoupled from $\phi$, both at the level of the Lagrangian and as far as their contributions to the stress-energy tensor are concerned. The systems is thus divided into two sectors: a transverse one, for which the presence of $\phi$ is irrelevant, and which thus has vanishing energy density and pressure in dimensional regularization,
\be
\rho_T =  \int \frac{d^d \textbf{{k}} }{(2\pi)^d}  \, c_T |\textbf{{k}}| = 0 \; , \qquad p_T = \frac1{d} \int \frac{d^d \textbf{{k}} }{(2\pi)^d} \,  c_T |\textbf{{k}}| = 0 \; ,
\ee
and a longitudinal one, consisting of $\vec \eta_L$ and $\phi$, for which the computation of the energy and pressure
is, as we will now see, quite involved. 

From now on, we will restrict to the longitudinal sector, which amounts to setting $\vec \eta_T$ to zero in the formulae above. Also, for notational simplicity we will drop the subscript `$L$' from $\vec \eta_L$ (but we will remember that we are dealing with a longitudinal field.)

%%%%%%%%%%%%%%%%%%%%%%%%%%%%%%%%%%%%%%%%%%%
%%%%%%%%%%%%%%%%%%%%%%%%%%%%%%%%%%%%%%%%%%%
\subsection{Analysis for $b_3$}\label{b3}
Consider first the case in which $b_1$ and $b_2$ are set to zero while $b_3$ is nonzero. In that case, $\vec \eta$ and $\phi$ are completely decoupled (at this order), and their contributions to $\rho + p$ can be analyzed separately. For $\vec \eta$, there isn't much to say: the relevant integrals for $\rho$ and $p$ vanish in dimensional regularization, because they do not involve any mass scale---that is, they are integrals of pure powers.

The situation is more interesting for $\phi$. Its action is that of a massive scalar with a  propagation speed
different from one,
\be \label{action b3}
S = \int d^4 x  \, \sfrac12  \big[ \dot \phi^2 -   c_\phi^2 (\vec \nabla \phi)^2  -   M_\phi^2 \phi^2  \big] \; , \qquad \qquad c^2_\phi \equiv \frac1{1+b_3} \; , \qquad M_\phi \equiv c_\phi m \; ,    
\ee
where for convenience we redefined the normalization of $\phi$ by $\phi \to c_\phi \phi$.

Applying to the purely $\phi$ parts of eqs.~\eqref{53}, \eqref{54} the same manipulations as in the case of a relativistic scalar (see sect.~\ref{section2}),
we find that the integrals we should compute are
\be
\rho_{\phi} = \frac12 \int \frac{d^d \textbf{{k}}}{(2\pi)^d} \sqrt{c^2_\phi \textbf{{k}}^2 + M^2_\phi}  \; , \qquad 
p_{\phi} =c^2_\phi \times  \frac1{2d} \int \frac{d^d \textbf{{k}}}{(2\pi)^d} \frac{\textbf{{k}}^2}{\sqrt{c^2_\phi \textbf{{k}}^2 + M^2_\phi}} \; .
\ee
Up to a redefinition of the integration variable, $\textbf{{k}} = \textbf{{k}} \, ' /c_\phi$, these are clearly the same integrals as those of sect.~\ref{section2}. What's perhaps surprising is that, after such a change of variables, the overall powers of $c_\phi$ we are left with are the same for $\rho_\phi$ and for $p_\phi$:
\be \label{rho and p of phi}
\rho_\phi = \frac{1}{c_\phi^d} \, \rho_{\rm rel}(M_\phi) \; , \qquad 
p_\phi = \frac{1}{c_\phi^d} \, p_{\rm rel} (M_\phi)\; ,
\ee
where $\rho_{\rm rel}(m)$ and $p_{\rm rel}(m)$ are the energy density and pressure for a relativistic scalar of mass $m$, eqs.~\eqref{rho dim reg} and \eqref{p dim reg}.
So, once again, we get
\be
\rho_\phi + p_\phi = 0 \; .
\ee

This apparent accident is in fact a consequence of a formal (spurionic) invariance of the action \eqref{action b3}: if we rescale the spatial coordinates but not time, and compensate for this by a rescaling of $c_\phi$ and of $\phi$, 
\be
\vec x \to \lambda \, \vec x \; , \qquad c_\phi \to \lambda \, c_\phi \; , \qquad \phi \to \lambda^{-3/2} \phi \; , 
\ee
the action does not change. As usual for spurion analyses, this informs how physical quantities can depend on $c_\phi$. In particular, we are interested in
\be
\rho_{\phi} = \langle T^{00} \rangle \; , \qquad p_{\phi}  = \sfrac1d \langle T^{ii} \rangle \; ,
\ee
for a state that is invariant under translations. These expectation values are thus constant in $\vec x$, and so their transformation properties under rescalings of coordinates must be completely taken care of by explicit powers of $c_\phi$. $T^{00}$ is a spatial density of energy. Since energy$\,\sim\,$time$^{-1}$ does not change under a rescaling of spatial coordinates, we must have
\be
T^{00} \to \frac{1}{\lambda^d} T^{00} \quad \Rightarrow \quad \langle T^{00} \rangle \propto \frac{1}{c_\phi^d} \; .
\ee
$T^{ij}$ is not the density of a conserved quantity. However, it is related to the momentum density $T^{0j}$ by the conservation equation
\be \label{conserve}
\partial_0 T^{0j} + \partial_i T^{ij} = 0 \; .
\ee
The momentum density rescales as
\be \label{T0j}
T^{0j} \to \frac{1}{\lambda^{d+1}} T^{0j} \; ,
\ee 
because momentum itself is the inverse of a length, and so it must rescale as $\vec P \to \vec P / \lambda$.
Using \eqref{conserve} and \eqref{T0j}, we thus have
\be
T^{ij} \to \frac{1}{\lambda^d} T^{ij} \quad \Rightarrow \quad \langle T^{ij} \rangle \propto \frac{1}{c_\phi^d}
\ee

We thus see that both $\rho_\phi$ and $p_\phi$ depend on $c_\phi$ exactly in the same way, in agreement with our explicit result in \eqref{rho and p of phi}. 

In more intuitive terms, all of the above stems from the statement that, even if we abandon natural units and we give independent units to mass ($m$), length ($\ell$), and time ($t$), energy density and pressure still have the same units\footnote{For instance, in SI units, 1 J/m$^3$ = 1 kg/m\,s$^2$ = 1 Pa.}:
\be
\Big[\frac{\mbox{energy}}{\mbox{volume}}\Big] = \frac{m}{\ell \cdot t^2 } \; , \qquad \big[\mbox{pressure}\big]  =\Big[\frac{\mbox{force}}{\mbox{surface}}\Big] = \frac{m}{\ell \cdot t^2 } \; .
\ee
Their ratio is thus dimensionless, and must be the same in all unit systems. In particular if it is $-1$ in units such that $c_\phi =  1$, then it must be $-1$ even when $c_\phi \neq 1$.  

The arguments above show that, as far as our check is concerned, we can consistently set $b_3$ to zero, even when we turn $b_1$ and $b_2$ back on. The reason is that we can work in units such that $c_{\phi} = 1$, which corresponds to $b_3 = 0$, and run our check in those units. Going from natural units ($c=1$) to $c_{\phi} = 1$ units certainly affects the values of the other parameters of the theory: $c_L$, $c_T$, $M_1$, $b_1$, $b_2$, and $m$. However, since we are leaving these generic anyway, such a change has no repercussions for our check\footnote{For more general questions, say computing $\rho$ and $p$ separately rather that just checking $\rho + p = 0$, one can still first work in $c_\phi = 1$ units and then reinstate the dependence on $c_\phi$ after the fact, by taking into account how the other parameters change when we change units. For instance the parameter $c_L$ in $c_\phi =1$ units becomes $c_L/c_\phi$ in any other unit system.}. %delete "after the fact"? (GS)

%%%%%%%%%%%%%%%%%%%%%%%%%%%%%%%%%%%%%%%%%%%
%%%%%%%%%%%%%%%%%%%%%%%%%%%%%%%%%%%%%%%%%%%
\subsection{Analysis for $b_1$ and $b_2$}\label{b1 and b2}

We may then switch off $b_3$ and perform our computation for non-zero values of the mixing coefficients $b_1$ and $b_2$. Combining eqs.~\eqref{53} and \eqref{54}, switching to a generic number of spatial dimensions, $3\to d$, and dropping again terms proportional to the quadratic Lagrangian, we obtain the following expression for $\rho + p$: 
\begin{align}
\rho+ p \equiv   \langle T^{00} + \frac{1}{d}T^{ii} \rangle &= \frac{d+1}{d}\left(\dot{\vec{\eta}} \,^2+\dot\phi^2\right) -\frac{1}{d}m^2\phi^2 +\frac{1}{d}b_1 \, \phi\,\vec\nabla\cdot\vec\eta + \frac{d+1}{d}b_2 \, \dot\phi\,\vec\nabla\cdot\vec\eta \; . \label{qq}
\end{align}
We can now perform the same manipulations as in sect.~\ref{section2}. We find:
\begin{align}
    \rho+p = & \; \frac{1}{d}\lim_{x\to y}\Big[-(d+1)\partial_t^2\langle\vec\eta(x)\cdot\vec\eta(y)\rangle -((d+1)\partial_t^2+m^2)\langle\phi(x)\phi(y)\rangle \nonumber\\
    &+(b_1-(d+1)b_2\partial_t)\vec\nabla\cdot\langle\vec\eta(x)\phi(y)\rangle \Big] \nonumber \\
    =& \; \frac{1}{d}\int \frac{d^{d}\textbf{{k}}}{(2\pi)^{d}}\frac{d\omega}{2\pi}\Big[ (d+1) \, \omega^2 \, \tilde G_W^{\eta\eta}(\omega, \textbf{{k}} \,) + ((d+1) \, \omega^2  - m^2) \, \tilde G_W^{\phi\phi}(\omega, \textbf{{k}} \, ) \nonumber\\
    &+( ib_1 -(d+1)b_2 \, \omega  )\, |\textbf{{k}}| \,  \tilde G_W^{\eta\phi}(\omega, \textbf{{k}} \,)\Big] \label{14} \; ,
\end{align}
where $\tilde G_W^{ab }$ is the matrix of Wightman two-point functions for the fields $\psi^a = (\eta, \phi)$ and, thanks to the longitudinality of $\vec \eta$, we  were able to switch to a purely scalar notation, %This is probably the only place where we do not put _W for a Wightman function... (GS)
\be
   \tilde {\vec \eta} (\omega, \textbf{{k}} \, )\equiv \hat k \, \tilde  \eta  (\omega, \textbf{{k}} \, ) \; .
\ee

As usual, from the quadratic Lagrangian we can easily compute the matrix of Feynman propagators, $\tilde {G}^{ab}_F(\omega, \textbf{{k}} \, )$. However, going from these to the Wightman two-point functions  requires some work. The general relationship, which we review in Appendix \ref{App A}, is
\begin{align}
\tilde {G}_W^{ab}(\omega, \textbf{{k}} \, ) & = \int \frac{d\omega'}{2\pi} \left[ \frac{i}{\omega - \omega' + i\epsilon} \tilde {G}_F^{ab} (\omega', \textbf{{k}} \,) -  \frac{i}{\omega - \omega' - i\epsilon} \tilde {G}^{ab \, ^*}_{F} (-\omega', - \textbf{{k}} \, )  \right] \label{contour} \\
& = \sum_n \left[ K^{-1}(\omega, \textbf{{k}} \,) (\omega - \omega_n) \right]^{ab} (2\pi)\delta(\omega - \omega_n) \; ,
\end{align}
where $K$ is the kinetic matrix appearing in the quadratic action,
\be
S = \frac{1}{2}\int \frac{d^3 \textbf{{k}}}{(2\pi)^3} \frac{d\omega}{2\pi} \,  \tilde \psi {}^{*a}(\omega,\textbf{{k}}\,) K_{ab}(\omega,\textbf{{k}} \, ) \tilde \psi^b(\omega, \textbf{{k}} \,)  \; , 
\ee
and the $\omega_n = \omega_n(\textbf{{k}} \,)$ are the positive-energy poles of the Feynman propagators.

Using this in eq.~\eqref{14} and performing the $\omega$ integral leaves us with
\begin{align} \label{rho+p residues}
\rho + p =  \frac{1}{d} \sum_n \int \frac{d^d \textbf{{k}}}{(2\pi)^{d}} \Big[ & (d+1) \, \omega_n^2 \, R_n^{\eta\eta}(\textbf{{k}} \,) + ((d+1) \, \omega_n^2  - m^2) \, R_n^{\phi\phi}(\textbf{{k}} \,) \nonumber \\
    & +( ib_1 -(d+1)b_2 \, \omega_n  )\, | \textbf{{k}} |\,  R_n^{\eta\phi}(\textbf{{k}} \, ) \Big] \; , 
\end{align}
where the $R$'s are the residues
\be \label{residues}
R_n^{ab} (\textbf{{k}} \,)= \lim_{\omega \to \omega_n} \left[ K^{-1}(\omega, \textbf{{k}} \,) (\omega - \omega_n) \right]^{ab}   \; .
\ee

The positions of the poles can be computed for generic $b_1$ and $b_2$, and so can the associated residues. However, the resulting expressions involve somewhat complicated double square root structures, which makes it impossible for us to perform the final integral in $\textbf{{k}}$ explicitly. To circumvent this problem, we can consider the small $b_1, b_2$ limit, and expand to the first nontrivial order in these couplings.

The kinetic matrix for $\eta$ and $\phi$ associated with the quadratic Lagrangian \eqref{52} is
 \begin{equation}
        K= 
        \begin{pmatrix}
        \omega^2-c_L^2 \textbf{{k}}^2& - (b_2  \omega +ib_1)|\textbf{{k}}|\\
       - (b_2  \omega -ib_1)|\textbf{{k}}| & \omega^2-\textbf{{k}}^2 - m^2 
        \end{pmatrix}
    \end{equation}
The poles of the Feynman propagators are the zeros of $\det K$, which to quadratic order in $b_1$, $b_2$ read
 \begin{align}
        \omega_1 & \simeq c_L |\textbf{{k}}|-\frac{ |\textbf{{k}}|\;\big(b_1^2+b_2^2c_L^2 \textbf{{k}}^2\big)}{2c_L\big((1-c_L^2) \textbf{{k}}^2+m^2\big)} \; , \nonumber \\
        \omega_2 & \simeq \sqrt{\textbf{{k}}^2+m^2}+\frac{\textbf{{k}}^2\big(b_1^2+b_2^2(\textbf{{k}}^2+m^2)\big)}{2\sqrt{\textbf{{k}}^2+m^2}\big((1-c_L^2)\textbf{{k}}^2+m^2\big)} \; .
        \end{align}
Upon inverting $K$, expanding the residues \eqref{residues} also to quadratic order, and plugging these expansions into our expression for $\rho+p$, eq.~\eqref{rho+p residues}, we end up with
    \begin{equation}
    \rho+p = \frac{1}{d} \sum_A C_A \int \frac{d^d \textbf{{k}}}{(2\pi)^d}\frac{ |\textbf{{k}}| \,^{\alpha_A}}{\big((1-c_L^2)\textbf{{k}}^2+m^2\big)^{\gamma_A}}\sqrt{\textbf{{k}}^2+m^2} \,^{\beta_A}, \label{19}
    \end{equation}
    where the $C_A$'s are suitable coefficients, and $\alpha_A$, $\beta_A$, $\gamma_A$ are powers that vary in combinations, yielding in total eleven structurally distinct terms, as outlined in Table \ref{tab1}.

\begin{table}[h!]
  \centering
  \begin{tabular}{|M{1cm}|M{1cm}|M{1cm}| M{6cm}|}
    \hline
    $\bm{\alpha}$ & $\bm{\beta}$ & $\bm{\gamma}$ & $\bm{C}$ \\  %Subscript _A? (GS)
    \hline \hline
    1 & 0 & 0 & $\frac{1}{2}c_L (d+1)$ \\ \hline
    2 & -1 & 0 & $\frac{1}{2}$ \\ \hline
   0 & 1 & 0 & $\frac{d}{2}$ \\ \hline
    6 & -1 & 2 & $\frac{b_2^2}{4}(1-c_L^2)(d+1)$ \\ \hline
   4 & -1 & 2 & $\frac{b_1^2}{4}(d-1) +\frac{b_2^2}{4}m^2(3+d(2-c_L^2))$ \\ \hline
     2 & -1 & 2 & $\frac{b_1^2}{4}m^2(d+2)+\frac{b_2^2}{4}m^4(d+2)$ \\ \hline
   5 & 0 & 2 & $-\frac{b_2^2}{4}c_L(1-c_L^2)(d+1) $ \\ \hline
   3 & 0 & 2 & $-\frac{b_1^2}{4c_L}(1-c_L^2)(d-1)-\frac{b_2^2}{4}c_Lm^2(d+3)$ \\ \hline
   1 & 0 & 2 & $-\frac{b_1^2}{4c_L}m^2(d+1) $ \\ \hline
      6 & -3 & 2 & $-\frac{b_1^2}{4}c_L^2(d-1)$ \\ \hline
   4 & -3 & 2 & $-\frac{b_1^2}{4}c_L^2m^2d $ \\ \hline
  \end{tabular}
  \caption{Coefficients $C_A$ for all the values of the powers $\alpha_A$, $ \beta_A$, $\gamma_A$ that appear in \eqref{19}.}\label{tab1}
\end{table}

Notice that the first three terms in Table \ref{tab1} are the contributions one gets from the free theories of the Goldstones and the massive scalar. Namely, the first term is
\begin{equation}
\rho+p \supset \frac{1}{2}\int \frac{d^d \textbf{{k}}}{(2\pi)^d} \left(c_L | \textbf{{k}} |+\frac{1}{d}c_L |\textbf{{k}} |\right) \equiv \rho_L + p_L,
\end{equation}
in agreement with equations \eqref{38} and \eqref{39} and, since it involves only pure powers of the momentum, it integrates to zero in dimensional regularization. The second and third terms are the pressure and energy densities of the free massive scalar,
\begin{equation}
\rho+p \supset \frac{1}{2}\int \frac{d^d \textbf{{k}}}{(2\pi)^d}\left(\frac{1}{d}\frac{\textbf{{k}}^2}{\sqrt{\textbf{{k}}^2+m^2}}+\sqrt{\textbf{{k}}^2+m^2}\right)
\equiv p_\phi + \rho_\phi \; ,
\end{equation}
whose sum also equals zero, as was shown in section \ref{section2}.

%\item 
For the rest of the terms, we switch to polar coordinates and perform the integral in the radial $k$-direction; the closed-form result for this type of integrals, dropping the coefficients $C_A$, is
\begin{align}
    \frac{1}{2} &m^{d+\alpha_j+\beta_j-2\gamma_j}
    \Bigg[  \frac{(1-c_L^2)^\frac{-d-\alpha_j}{2}\Gamma\big(\frac{d+\alpha_j}{2}\big)\Gamma\big(\frac{-d-\alpha_j+2\gamma_j}{2}\big)\;{}_2F_1\big(\frac{d+\alpha_j}{2},-\frac{\beta_j}{2};\frac{2+d+\alpha_j-2\gamma_j}{2};\frac{1}{1-c_L^2}\big)}{\Gamma(\gamma_j)}   \nonumber \\
    &+ \frac{(1-c_L^2)^{-\gamma_j}\Gamma\big(\frac{d+\alpha_j-2\gamma_j}{2}\big)\Gamma\big(\frac{-d-\alpha_j-\beta_j+2\gamma_j}{2}\big)\; _2F_1\big(\gamma_j,\frac{-d-\alpha_j-\beta_j+2\gamma_j}{2};\frac{2-d-\alpha_j+2\gamma_j}{2};\frac{1}{1-c_L^2}\big)}{\Gamma(-\frac{\beta_j}{2})}\Bigg], \label{20}
    \end{align}   
where $_2F_1$ is a hypergeometric function. For some structures in the integrand of equation \eqref{19}, the integration result takes much simpler forms; for instance the term with $\alpha=5,\beta=0,\gamma=2$ yields
        \begin{equation}
            \int\frac{d^d \textbf{{k}}}{(2\pi)^d}\frac{|\textbf{{k}}| ^5}{\big((1-c_L^2) \textbf{{k}}^2+m^2 \big)^2}=-\frac{\pi}{4}m^{1+d}(1-c_L^2)^{\frac{-d-5}{2}}(3+d)\sec\Big(\frac{d\pi}{2} \Big).
        \end{equation}
    Other structures, particularly when $\beta\neq0$, integrate to totally nontrivial combinations of hypergeometric, Gamma, and trigonometric functions, as seen in the generalized form in \eqref{20}, and so displaying them here wouldn't provide much intuition.  
%    \end{itemize}
It is miraculous to see that all these terms, when summed, cancel exactly with each other, despite the highly complicated structures and combinations of coefficients involved. What's also interesting, is that we don't need to specify a number of spatial dimensions to obtain the final answer. That is, for any $d$, we find:
\begin{equation}
     \rho + p =0 \; .
\end{equation}

%%%%%%%%%%%%%%%%%%%%%%%%%%%%%%%%%%%%%%%%%%%
%%%%%%%%%%%%%%%%%%%%%%%%%%%%%%%%%%%%%%%%%%%

\section{Superfluid check}\label{superfluid check}

As a check of our methods and computational tools, we now look at the case of a superfluid, for which we know that the stress-energy tensor does {\em not} have a Lorentz invariant expectation value. In particular, we want to make sure that the mysterious cancellations that yield $\rho + p= 0$ in the framid case are not a result of potential nuances of the computation. For a superfluid,  $\langle T_{\mu\nu}(x)\rangle \neq -\Lambda \,\eta_{\mu\nu}$ already at tree level, and so an analogous computation should yield a nonzero result. 

The simplest implementation of a superfluid EFT involves a single scalar field $\psi(x)$ with a shift symmetry, $\psi \to \psi + {\rm const}$, and a time-dependent vev, $\langle\psi(x)\rangle=\mu t$ \cite{Son, NP, framid}, where $\mu$ is the chemical potential. The superfluid phonon field, $\pi(x)$, parametrizes fluctuations around this background, $\psi(x) \equiv \mu t + \pi(x)$. 
To lowest order in derivatives, the most general low-energy effective action is 
\begin{equation}
    S=\int d^4x \, P(X),\quad \quad X \equiv - \partial_\mu\psi\partial^\mu\psi,
\end{equation}
where $P(X)$ is a generic function, in one-to-one correspondence with the superfluid's equation of state \footnote{Namely, the function $P$ relates the pressure to the chemical potential: $p = P(\mu^2)$.}. Upon expanding to quadratic order in the phonon field and choosing canonical normalization, one gets
\begin{equation}
    S  \to  \frac{1}{2}\int d^4x\;\left[ \dot\pi^2 - c_s^2(\vec\nabla\pi)^2 \right] \; ,
\end{equation}
with the sound speed given in terms of derivatives of the Lagrangian,
\begin{equation}
    c_s^2 = \frac{P'(X)}{P'(X) + 2XP''(X)} \; .
\end{equation}
Similarly to the framid case, in order to run a nontrivial check in dimensional regularization we couple the superfluid to a massive scalar $\phi$ . We look for all possible shift-symmetric, Lorentz-invariant couplings that, when expanded in the superfluid phonons $\pi(x)$, yield quadratic terms in $\pi$ and $\phi$, with at most two derivatives acting on them. Since $X =  \mu^2+2\mu\dot\pi+\dot\pi^2-(\vec\nabla\pi)^2$, the  couplings that produce Lorentz-violating structures are: 
\begin{alignat*}{2}
& f_1(X)\phi &&\to \quad\dot\pi\phi \\
& f_2(X)\partial_\mu\psi\partial^\mu\phi &&\to \quad \dot\pi\dot\phi,\; \vec\nabla\pi\cdot\vec\nabla\phi \\
    &f_3(X)\partial_\mu\psi\partial_\nu\psi\partial^\mu\phi\partial^\nu\phi &&\to \quad \dot\phi^2 \; .
\end{alignat*}
For simplicity, we focus only on the first type of coupling, which introduces a $\dot\pi\phi$ term in the quadratic Lagrangian. Moreover, we choose the simplest possible form for $f_1$, that is $f_1(X) = X$. Finally, we work in the $c_s^2=1$ limit (which implies $P''(X)=0$). These choices give us the opportunity to compare our result directly to an independent path-integral calculation, which will be summarized below and published elsewhere \cite{JNPS}. 

So, in summary, we start from
\begin{equation}
    \mathcal{L} = P(X) + X\phi -\frac{1}{2}\partial_\mu\phi\partial^\mu\phi - \frac{1}{2}m^2\phi^2,
\end{equation}
whose  stress-energy tensor is
\begin{equation}
    T^{\mu\nu}=2P'(X)\partial^\mu\psi\partial^\nu\psi + 2\partial^\mu\psi\partial^\nu\psi\phi + \partial^\mu\phi\partial^\nu\phi,
\end{equation}
omitting, as before, terms proportional to the Lagrangian. Expanding to quadratic order and canonically normalizing $\pi$, we get
\begin{equation}
    \mathcal{L} \simeq \frac{1}{2}\left[\dot\pi^2-(\vec\nabla\pi)^2 + 4b \, \dot\pi\phi+ \partial_\mu\phi\partial^\mu\phi - m^2\phi^2\right] \; , \label{L2 superfluid}
\end{equation}
and 
\be
    T^{00} \simeq \dot\pi^2+\dot\phi^2 + 4b \, \dot\pi\phi \; , \qquad
    T^{ii} \simeq (\vec\nabla\pi)^2 + (\vec\nabla\phi)^2 \; , \label{Tmn superfluid}
\ee
where the spatial indices are summed over and $b\equiv \mu/\sqrt{2P'(\mu^2)} \,$ \footnote{Notice that the expressions \eqref{Tmn superfluid} can be obtained by  applying Noether's theorem directly to the quadratic Lagrangian \eqref{L2 superfluid}, but only if one takes into account that $\pi$ transforms nonlinearly under time-translations, since these are spontaneously broken by the background $\psi = \mu t$. This is related to the statement that the superfluid's ground state is an eigenstate of $H - \mu Q$, but not of $H$, with $H = \int d^3 x \,  T^{00}$ being the Hamiltonian, and $Q = \int d^3 x \,  J^0$ the charge.}. %I changed eqref{L2} to eqref{L2 superfluid} (GS)

Performing manipulations similar to those of the framid case, we rearrange some terms and obtain \begin{equation}
    T^{00}+\frac{1}{d}T^{ii}=\frac{d+1}{d}(\dot\pi^2+\dot\phi^2)-\frac{1}{d}m^2\phi^2 + 4\frac{d+1}{d}b \, \dot\pi\phi - \frac{2}{d}\mathcal{L} \; , 
\end{equation}
whose expectation value is given by
\begin{align}
\rho + p \equiv \langle T^{00}+\frac{1}{d}T^{ii}\rangle = \frac{1}{d}\int\frac{d^d \textbf{{k}}}{(2\pi)^{d}}\frac{d\omega}{2\pi}&\bigg[(d+1)\omega^2\left(\tilde G_W^{\pi\pi}(\omega, \textbf{{k}} \,)+\tilde G_W^{\phi\phi}(\omega, \textbf{{k}} \,)\right) - m^2 \tilde G_W^{\phi\phi}(\omega, \textbf{{k}} \,) \nonumber \\
&-4i(d+1)\omega b\, \tilde G_W^{\pi\phi}(\omega,  \textbf{{k}} \,)\bigg].
\end{align}
Like before, we expand the Feynman propagators and their poles for small values of the coupling constant $b$, and then construct the Wightman two-point functions. As expected, all expressions take simpler forms in this case. Namely, the (positive-frequency) poles are
\begin{align}
    \omega_1 & \simeq \sqrt{\textbf{{k}}^2+m^2} + \frac{2b^2}{m^2}\sqrt{\textbf{{k}}^2+m^2} \; , \nonumber \\
    \omega_2 & \simeq | \textbf{{k}}| -\frac{2b^2}{m^2}| \textbf{{k}}| \; ,
\end{align}
and the corresponding residues  are
\begin{alignat}{2}
   R_1^{\pi\pi}(\textbf{{k}} \,) &\simeq \frac{2b^2}{m^4}\sqrt{\textbf{{k}}^2+m^2} \; ,  && R_2^{\pi\pi}(\textbf{{k}} \,) \simeq  \left[\frac{1}{2|\textbf{{k}}|}-\frac{b^2(2\textbf{{k}}^2+m^2)}{m^4|\textbf{{k}}|}\right] \; ,\nonumber \\
    R_1^{\phi\phi}(\textbf{{k}} \,)& \simeq \left[\frac{1}{2\sqrt{\textbf{{k}}^2+m^2}}-\frac{b^2(2\textbf{{k}}^2+m^2)}{m^4\sqrt{\textbf{{k}}^2+m^2}}\right] \; , \quad \quad && R_2^{\phi\phi}(\textbf{{k}} \,) \simeq \frac{2b^2}{m^4}|\textbf{{k}}|\; , \nonumber \\
    R_1^{\pi\phi}( \textbf{{k}} \,)& \simeq -i\frac{b}{m^2} \; ,  && R_2^{\pi\phi}( \textbf{{k}} \,) \simeq i\frac{b}{m^2} \; .
\end{alignat}
Putting everything together, the integral in question eventually becomes
\begin{align}
\rho+ p & =\frac{b^2}{d \, m^2}\int\frac{d^d \textbf{{k}}}{(2\pi)^d}\frac{\textbf{{k}}^2-d(\textbf{{k}}^2+m^2)}{\sqrt{\textbf{{k}}^2+m^2}} \\
& = -\frac{b^2 m^2}{4\pi^2(d+1)}\left(\frac{m^2}{4\pi}\right)^{(d-3)/2}\Gamma \left(\frac{1-d}{2}\right) \; ,
\end{align}
which is different from zero, as expected.

We now come to the result of the independent computation alluded to above. As a possible UV completion of a superfluid EFT, one can consider a massive complex scalar $\Phi$ with quartic interactions. Putting the system at finite chemical potential, one ends up with a superfluid, with our field $\psi$ being associated with the phase of $\Phi$. On the other hand, the radial mode of $\Phi$ is massive, and can thus be thought of as our massive scalar $\phi$. For this system, one can explicitly compute first the associated $P(X)$ at tree level, and then the one-loop quantum corrections to it, in the form of the quantum effective action $\Gamma[\psi]$ \cite{JNPS}. (To lowest order in derivatives, this can be done via functional methods akin to those normally used to compute a Coleman-Weinberg effective potential \cite{CW, Weinberg}.) Applying  Noether's theorem to $\Gamma[\psi]$, one obtains directly the one-loop {\em expectation value} of $T^{\mu\nu}$, which, for $\rho+p$, matches exactly the result above \cite{JNPS}.

%%%%%%%%%%%%%%%%%%%%%%%%%%%%%%%%%%%%%%%%%%%
%%%%%%%%%%%%%%%%%%%%%%%%%%%%%%%%%%%%%%%%%%%

\section{Discussion}
We close with a few remarks:

\begin{enumerate}
\item
Our paper is about Lorentz-invariance in a system that spontaneously breaks it—that is, in a system in which such a symmetry is not manifest. So, one must be particularly careful in unveiling possible sources of Lorentz breaking coming from the way one does computations. We already discussed at length how to address UV divergences in a way that is compatible with Lorentz invariance. Another subtlety one needs to address is the Lorentz invariance of the path-integral measure.
Even though we  did not use path integrals for our computations, canonical quantization of the Goldstones' effective theory is equivalent to path-integral quantization with the measure
\be
D \vec \eta \equiv \prod_x d^3 \eta(x) \; .
\ee
This is not Lorentz invariant, because the $\vec \eta$ fields transform non-linearly under boosts
\footnote{Equivalently, one can phrase the problem directly in the canonical formalism \cite{Weinberg1}. See for instance ref.~\cite{ChiralLoops} for an analysis of the same issue for the chiral Lagrangian.}. As usual, this should not matter as long as one uses dimensional regularization, but with other regulators it might matter. We investigate the issue in Appendix \ref{measure}. The conclusion is that this subtlety does not matter:
({\em i}\,) for our quantity ($\langle T^{\mu\nu} \rangle$), to the order we are doing computations (one loop), regardless of the regulator used; or
({\em ii}\,) for any quantity, to any order, if one uses dimensional regularization. We were thus justified to neglect it.

\item
We have been emphasizing, puzzling over, and checking the fact that our expectation value $\langle T^{\mu\nu} \rangle \propto \eta^{\mu\nu}$ is Lorentz-invariant. However, as we hope the discussion in sect.~\ref{b3} has made clear, such an expectation value is in fact compatible with {\em any} Lorentz invariance, that is, it is invariant under generalized boosts with an arbitrary speed-of-light parameter $c$. This is because $\langle T^{00} \rangle$ (energy density) and $\langle T^{ij} \rangle$ (pressure, or stresses) always have the same units, and so a statement like $\langle T^{\mu\nu} \rangle \propto \eta^{\mu\nu} = {\rm diag}(-1,1,1,1)$ is independent of the units used. And so, in particular, it is independent of the value of $c$. We find this to be an interesting twist. It could have important conceptual implications, or it could just be a technical curiosity.

\item
Besides the framid, there are other cases in which the expectation value of $T^{\mu\nu}$ is more symmetric than the ground state \footnote{We thank Lam Hui for bringing this up.}. 
For instance, for a superfluid time-translations are spontaneously broken, but the expectation value in question is invariant under them (see sect.~\ref{superfluid check}).
However, in that case such a symmetry property can be explained by standard selection rules: the ground state spontaneously breaks time translations ($H$) and a $U(1)$ symmetry ($Q$) down to a linear combination thereof ($H - \mu Q$) \cite{NP}. Expectation values and more in general correlation functions should only be invariant under the unbroken combination. However, for operators that are neutral under the $U(1)$ symmetry, this automatically  translates into invariance under time-translations. $T^{\mu\nu}$ is one such operator, and so its expectation value is invariant under time translations, even though the ground state is not. We do not see any mechanism like this at play in the framid case.

\end{enumerate}

\noindent
What is the general lesson of our analysis? Are there implications for the cosmological constant problem? 
In general terms, our analysis exhibits an explicit example of a quantum system in which a certain expectation value is invariant under a symmetry even though there are no selection rules (including those of remark 3 above) enforcing this. In order to find a potential application of this phenomenon to the cosmological constant problem, we think one should start by making progress in two directions. The first is to understand how general this phenomenon is: are there other examples, and what are their common features---for instance, do they all require a spontaneous breaking of Lorentz invariance? The second is to find a structure, a pattern in our one-loop check: the cancellations that lead to $\rho + p = 0$ for the framid, especially those of sect.~\ref{b1 and b2}, are absolutely nontrivial. It is hard to believe that they are not enforced by a hidden structure in the computation. Perhaps there is a better way of organizing the computation that would make such a structure manifest. We plan to explore these questions in the near future.

%%%%%%%%%%%%%%%%%%%%%%%%%%%%%%%%%%%%%%%%%%%
%%%%%%%%%%%%%%%%%%%%%%%%%%%%%%%%%%%%%%%%%%%
\section*{Acknowledgements}
We thank Lam Hui, Riccardo Penco, and Raman Sundrum for useful discussions and comments. Our work is partially supported by the US DOE (award number DE-SC011941) and by the Simons Foundation (award number 658906).

%%%%%%%%%%%%%%%%%%%%%%%%%%%%%%%%%%%%%%%%%%%
%%%%%%%%%%%%%%%%%%%%%%%%%%%%%%%%%%%%%%%%%%%
\appendix
\section{Wightman and Feynman}\label{App A}
As we have seen, it is particularly helpful to rewrite the expectation value of the stress-energy tensor in terms of  derivatives acting on Wightman two-point functions. Formally, given a set of real fields $\psi^a$ governed by a quadratic Lagrangian,  in general one has
\be{\label{tmninwightman}}
\langle T^{\mu \nu} (0)\rangle = \lim_{x \rightarrow 0} \sum_{ab} D_{ab}^{\mu\nu}(\partial_x) G^{ab}_W(x) \; ,
\ee
where $D_{ab}^{\mu\nu}(\partial_x)$ is a function of derivatives, usually up to second order, and $G_W$ represents the matrix of Wightman two-point functions,
\be
G^{ab}_W(x) \equiv \langle \psi^a(x) \psi^b(0) \rangle \; .
\ee
Notice that, for real fields, one has
\be
G^{ba}_W(x) = G^{ab \, *}_W(-x) \; .  \label{GW real}
\ee

However, we are more familiar with the calculation of Feynman propagators given a certain theory, not the Wightman version. How can we relate $G_W$ to $G_F$, the Feynman propagator, in a way that is helpful to our calculations? 

To begin with, notice that, by definition,
\begin{align}
G^{ab}_F & = \langle \psi^a(x) \psi^b(0) \rangle \theta (t) + \langle \psi^b(0) \psi^a(x) \rangle \theta (-t) \label{GF define} \\
& = G^{ab}_W(x)\theta(t) + G^{ba}_W(-x)\theta(-t) \\
& = G^{ab}_W(x)\theta(t) + G^{ab \, *}_W(x)\theta(-t) \; .
\end{align}
As usual, $\theta(t)$ is the step function. This relation can be inverted to give 
\be
G^{ab}_W(x) = G^{ab}_F(x)\theta(t) + G^{ab \, *}_{F}(x)\theta(-t) \; .
\ee

Computations are usually easier in Fourier transform. Using the Fourier representation of the step function,
\be
\theta (t) = \int \frac{d \omega}{2 \pi} \frac{i}{\omega+i \epsilon} e^{-i \omega t} \; , 
\ee
we get
\begin{align}
{\label{gwingf}}
\tilde{G}_W^{ab}(\omega, \textbf{{k}} \, ) & = \int \frac{d\omega'}{2\pi} \left[ \frac{i}{\omega - \omega' + i\epsilon} \tilde{G}_F^{ab} (\omega', \textbf{{k}} \,) -  \frac{i}{\omega - \omega' - i\epsilon} \tilde{G}_F^{ab \, *} (-\omega', -\textbf{{k}} \,) \right] \\
& =\int \frac{d\omega'}{2\pi} \left[ \frac{i}{\omega - \omega' + i\epsilon} \tilde{G}_F^{ab} (\omega', \textbf{{k}} \,) + {\rm h.c.} \right] \;, \label{GW integral}
\end{align}
where the last equality follows from $\tilde{G}^{ab}_F(\omega,\textbf{{k}} \,) = \tilde{G}^{ba}_F(-\omega, -\textbf{{k}} \,)$---a direct consequence of the definition \eqref{GF define}.
We thus see that $\tilde{G}^{ab}_W$ is a hermitian matrix, in agreement with \eqref{GW real}. 

The matrix of Feynman propagators is easily computed starting from the quadratic action written in Fourier space,
\be
S = \frac{1}{2}\int \frac{d^3 \textbf{{k}}}{(2\pi)^3} \frac{d\omega}{2\pi} \tilde \psi^{*a}(\omega,\textbf{{k}} \,) K_{ab}(\omega,\textbf{{k}} \, ) \tilde \psi^b(\omega, \textbf{{k}} \,) \label{K mat} \; ,
\ee
where the kinetic matrix $K$ is hermitian.
In matrix notation, one simply has
\be
\tilde{G}_F (\omega, \textbf{{k}} \,) = i \big(K(\omega,\textbf{{k}} \,) + i \epsilon \big)^{-1}  \; .
\ee

Focusing on the first term in the integral \eqref{GW integral}, and assuming that, as usual, the Feynman propagators decay at infinity and have (simple) poles slightly away from the real axis, we can close the $\omega'$ contour in the lower half plane. We thus only pick up the poles of $\tilde{G}_F$ that lie under the real axis---the positive frequency ones, for a stable theory. We get
\be
\tilde{G}_W^{ab}(\omega, \textbf{{k}} \, ) = \sum_n \frac{i}{\omega - \omega_n + i\epsilon}  \left[ K^{-1}(\omega', \textbf{{k}} \,) (\omega' - \omega_n) \right]_{\omega' = \omega_n} ^{ab} + {\rm h.c.} \; ,
\ee
where the sum is extended over the positive frequency poles, $\omega_n = \omega_n(\textbf{{k}} \,)$.

Using the distributional identity
\be
\frac{1}{x+i\epsilon} = P\frac1x - i \pi \delta(x)
\ee
and the hermiticity of $K$, we finally get
\be
\tilde{G}_W^{ab}(\omega, \textbf{{k}} \, ) = \sum_n  \left[ K^{-1}(\omega', \textbf{{k}} \,) (\omega' - \omega_n) \right]_{\omega' = \omega_n} ^{ab} \, (2\pi) \delta(\omega - \omega_n) \; .
\ee

As a check, for a single relativistic massive scalar  the kinetic ``matrix'' is simply
\be
K = \omega^2 - \textbf{{k}} \, ^2 - m^2  \; , 
\ee
the positive frequency pole is
\be
\omega_k = \sqrt{\textbf{{k}} \, ^2 + m^2} \; ,
\ee
and the Wightman two-point function thus reduces to
\begin{align}
\tilde G_W (\omega, \textbf{{k}}) & = \frac{1}{2 \omega_k} (2 \pi) \delta(\omega - \omega_k)\\
& = (2\pi) \theta(\omega) \delta(k^2+m^2) \; ,
\end{align}
which is the correct expression. Notice however that, in the general derivation above, we have never used Lorentz invariance.

%%%%%%%%%%%%%%%%%%%%%%%%%%%%%%%%%%%%%%%

\section{Symmetric stress-energy tensors}
One may consider performing our calculations starting from the more trusted symmetric versions of the stress-energy tensor, i.e.~the Hilbert and Belinfante tensors. When deriving the Hilbert tensor for a framid, we have to keep in mind the unit-norm constraint on $A^\m$,
\begin{equation}
g^{\m\n}A_\m A_\n = -1,
\end{equation}
which forbids varying the metric $g^{\m\n}$ independently of $A^\m$. One can introduce a vierbein and vary the action with respect to it instead, yielding \cite{framid}
\begin{equation}
T_H^{\m\n} = \frac{1}{\sqrt{-g}}\left(2\frac{\delta S}{\delta g_{\m\n}}+\frac{\delta S}{\delta A_\m}A^\n \right) \; ,
\end{equation}
where now the functional derivatives are unconstrained.

The tensor above is evidently not  symmetric in general. In fact, it is symmetric only on-shell, i.e~upon using the equations of motion, which, taking into account the unit-norm constraint once more, are simply
\be
(\eta_{\mu\nu} + A_\mu A_\nu) \frac{\delta S}{\delta A_\nu} = 0 \; .
\ee
The end result for the (symmetric) Hilbert tensor is
\begin{align}
T_H^{\m\n} &=\mathcal{L}g^{\m\n} \nonumber\\
&+ 2c_1 \Big[  A^{(\m} \partial_{\alpha}\partial^{\n)}A^{\alpha} -  A_{\alpha} \partial^{\alpha}\partial^{(\m}A^{\n)}   -  \partial^{\alpha}A_{\alpha} \partial^{(\m}A^{\n)}  -  \partial^{(\m}A^{\alpha} \partial^{\n)}A_{\alpha}  + \partial_{\alpha}A^{(\m} \partial^{\n)}A^{\alpha}  \Big] \nonumber\\
&+2c_2 \Big[A^{(\m} \partial_{\alpha}\partial^{\n)}A^{\alpha}   -  g^{\m\n} A_{\alpha}  \partial_{\beta}\partial^{\alpha}A^{\beta} -  g^{\m\n} \partial_{\alpha}A^{\alpha} \partial^{\beta}A_{\beta}\Big] \nonumber\\
&+ 2c_3 \Big[ A^{(\m} \partial_{\alpha}\partial^{\alpha}A^{\n)} + \partial_{\alpha}A^{(\m} \partial^{\alpha}A^{\n)} -  A_{\alpha} \partial^{\alpha}\partial^{(\m}A^{\n)}  -  \partial^{\alpha}A_{\alpha} \partial^{(\m}A^{\n)} -  \partial_{\alpha}A^{(\m} \partial^{\n)}A^{\alpha} \Big] \nonumber \\
&+ 2c_4 \Big[ A^{\m} A^{\n} \partial_{\alpha}A^{\beta} \partial_{\beta}A^{\alpha} + A^{\alpha} A^{\m} A^{\n} \partial_{\beta}\partial_{\alpha}A^{\beta} -  A^{\alpha} A^{(\m} \partial_{\alpha}A^{\n)} \partial^{\beta}A_{\beta} -  A^{\alpha} A_{\beta} \partial_{\alpha}A^{(\m} \partial^{\beta}A^{\n)}   \nonumber\\
& \quad\quad -  A^{\alpha} A_{\beta} A^{(\m} \partial^{\beta}\partial_{\alpha}A^{\n)} + A^{\alpha} A^{(\m} \partial_{\alpha}A^{\beta} \partial_{\beta}A^{\n)}   -  A^{\alpha} A^{(\m} \partial_{\alpha}A_{\beta} \partial^{\n)}A^{\beta} \Big] \; ,\label{hilb}
\end{align}
where the $c_a$'s are coefficients related to the $M_1^2$, $c_L^2$, and $c_T^2$ coefficients of sect.~\ref{framid} \cite{framid}.
Manipulating this tensor to compute our quantum corrections clearly requires considerably more effort compared to the Noether one, eq.~\eqref{8}. The same is true for the Belinfante tensor, which turns out to be exactly the same as the Hilbert one.

Namely, the Belinfante stress-energy tensor in general is defined   as \cite{Weinberg1}
\begin{equation}
T^{\m\n}_B = T^{\m\n}_N -\frac{i}{2}\del_\kappa\left[\frac{\del\mathcal{L}}{\del(\del_\kappa \Phi^a)}(\mathcal{J}^{\m\n})^a {}_b \Phi^b -\frac{\del\mathcal{L}}{\del(\del_\m \Phi^a)}(\mathcal{J}^{\kappa\n})^a {}_b \Phi^b - \frac{\del\mathcal{L}}{\del(\del_\n \Phi^a)}(\mathcal{J}^{\kappa\m})^a {}_b \Phi^b \right], \label{15}
\end{equation}
where $T^{\m\n}_N$ is the Noether stress-energy tensor, and the $\mathcal{J}^{\m\n}$'s are the Lorentz generators in the representation appropriate for the fields $\Phi^a$. For 4-vector fields,
\begin{equation}
(\mathcal{J}^{\rho\sigma})^\kappa {}_\lambda = i(\h^{\sigma\kappa}\delta^\rho_\lambda - \h^{\rho\kappa}\delta^\sigma_\lambda) \; .
\end{equation}
Writing down all terms in \eqref{15}, the result is equal to equation \eqref{hilb} plus terms that are proportional to the equations of motion\footnote{Note that the Belinfante tensor is also non-symmetric unless one uses the equations of motion  \cite{Weinberg1}.}. 
We checked that upon expanding the Belinfante and Hilbert stress-energy tensors to quadratic order in our $\vec \eta$ fields, we get exactly the same expressions for $\langle T^{00} \rangle$ and $\langle T^{ij} \rangle$ as those derived from the Noether stress-energy tensor, eqs.~\eqref{rho framid}, \eqref{p framid}.

%We reason then, that using either of these two symmetric tensors is unnecessary because 
%\begin{equation}
%T^{\m\n}_H = T_B^{\m\n}\quad \textup{(+ terms}\propto \textup{ EOM)} = T^{\m\n}_c + \partial [...],
%\end{equation}
%
%and the last term is going to vanish when evaluated on the ground state, since spacetime-translations are unbroken and
%\begin{equation}
%\langle \partial[...] \rangle = \partial \langle [...]\rangle = 0.
%\end{equation}
%Therefore, for evaluating the vev of the framid stress-energy tensor, which is what we want to do, all different forms of the  tensor are completely equivalent and can be reduced to $\langle T_c^{\m\n} \rangle$. Hence we choose to work with the simpler one - the canonical stress-energy tensor, as derived by Noether's theorem.
%

Notice that adding terms proportional to the equations of motion is one of the ambiguities inherent in the definition of the stress-energy tensor, or, for that matter, of any Noether current. Consider in fact the standard Noether procedure to derive a conserved current. It starts with inspecting how an action changes under a symmetry transformation if the transformation parameter  $\epsilon$ of a global symmetry of the form
\be
\Phi^a \to \Phi^a + \epsilon \Delta^a[\Phi] 
\ee
(for some functional $\Delta^a$)
is modulated weakly in space and time. But this prescription is ambiguous. The standard way to implement it is
\be
\Phi^a \to \Phi^a + \epsilon(x) \Delta^a[\Phi]  \; ,
\ee
but an equally valid one is, for example, 
\be
\Phi^a \to \Phi^a + \epsilon(x) \Delta^a[\Phi] + \partial^\mu \epsilon(x) \, F_\mu ^a[\Phi] \; ,
\ee
for an arbitrary functional $F^a$. These two approaches both yield conserved currents, and the two currents differ by a term proportional to the equations of motion, $ \frac{\delta S}{\delta \Phi^a} F_\mu^a[\Phi]$.

%%%%%%%%%%%%%%%%%%%%%%%%%%%%%%
%%%%%%%%%%%%%%%%%%%%%%%%%%%%%%
\section{The path-integral measure}\label{measure}
%Our paper is about Lorentz-invariance in a system that spontaneously breaks it---that is, in a system in which such a symmetry is not manifest. So, we must be particularly careful in unveiling possible sources of Lorentz breaking coming from the way we do computations. We already mentioned in the introduction how we plan to address UV divergences in a way that is compatible with Lorentz invariance. Another subtlety we need to address is the Lorentz invariance of the path-integral measure.
%
%Even though we will not be using  path integrals for our computations, canonical quantization of the Goldstones' effective theory is equivalent to path-integral quantization with the measure
%\be
%D \vec \eta \equiv \prod_x d^3 \eta(x) \; .
%\ee
%This is not Lorentz invariant, because the $\vec \eta$ fields transform non-linearly under Lorentz. As usual, this should not matter as long as we use dimensional regularization, but with other regulators it might matter. Let's thus investigate the issue further. 

In order to construct a Lorentz-invariant measure, we can start from the obvious invariant measure for $A_\mu$, 
\be
D A_{\mu} \equiv  \prod_x d^4 \! A(x) \; ,
\ee
and impose an invariant constraint that removes its norm, e.g.
\be
\delta(A_\mu A^\mu + 1) \; .
\ee
We can then parametrize $A_\mu$ in terms of our Goldstone fields $\vec \eta(x)$ and of a radial mode
$\rho(x)$,
\be
A_0 = \rho \, \cosh |\vec \eta \, | \; , \qquad \vec A = \rho \, \frac {\vec \eta}{ |\vec \eta \, |}  \sinh  |\vec \eta \, | \; .
\ee
The path integral then reads 
\be \label{PI}
\int D A_{\mu} \, \delta(A_\mu A^\mu + 1)\,   e^{i S} \dots = \int D \rho \, D \vec \eta \;  {\rm Det} J \; \delta(\rho^2- 1)\,   e^{i S} \cdots \; ,
\ee
where the dots denote insertions of operators, and the functional Jacobian $J$ is
\be
J(x, x') \equiv \frac{\delta \big(A_0(x), \vec A(x) \big)}{\delta \big(\rho(x'), \vec \eta(x') \big)} = \frac{\partial \big(A_0, \vec A \, \big)}{\partial \big(\rho, \vec \eta \,  \big)} \delta(x-x') \; .
\ee
Using standard functional methods \cite{Weinberg}, its determinant can be written in exponential form as
\be \label{det J}
 {\rm Det} J  = e^{i \Delta S} \; , \qquad \Delta S \equiv  -i \Big(\int \frac{d^4 k}{(2\pi)^4} \Big) \int d^4 x \, \log \frac{\sinh ^2 |\vec \eta \, | }{|\vec \eta \, |^2} \; ,
\ee
where we used that, thanks to the delta-function in \eqref{PI}, $\rho = 1$.
The integral over $\rho$ can now be performed explicitly, upon which we are left with the path integral
\be
\int  D \vec \eta \;   e^{i (S + \Delta S)} \cdots \; .
\ee
We thus reach the conclusion that, to preserve Lorentz invariance in our computations, we should  supplement the $\vec \eta$ effective action with  $\Delta S$. 

If we use dimensional regularization, $\Delta S$ vanishes, because its overall coefficient does. This is one of the many reasons why dimensional regularization is  convenient, and why we usually don't track functional determinants coming from field redefinitions in the path integral. 

If, on the other hand, we use other UV regulators, we should keep $\Delta S$ around. Notice that, like all effects coming from functional determinants, $\Delta S$ is formally of one-loop order. We should then use it consistently in perturbation theory. For instance, for one-loop computations, we should use $\Delta S$ at tree-level. 

$\Delta S$ is a (UV divergent) potential for our Goldstone fields. In particular, it includes a mass term for them. This is inconsistent with the Goldstone theorem for spontaneously broken boosts \cite{AN}. This means that, at one-loop, there must be other contributions that cancel at least the effects of such a mass term. Or, conversely, if we don't keep $\Delta S$ around, at one-loop we must find nontrivial contributions to the mass of the Goldstones, in violation of the Goldstone theorem.

 We can  check this explicitly. For simplicity, let's consider the $c_L = c_T = 1$ case, which is particularly symmetric \cite{framid}. The two-derivative Goldstone action takes the form of a relativistic non-linear sigma model, 
\be
S  = - \frac{M_1^2}{2} \int d^4 x f_{ij}(\vec \eta \, ) \partial_\mu \eta^i \partial^\mu \eta^j \; ,
\ee 
with $f_{ij}$ given by
\be
f_{ij} (\vec \eta \,) = P_{ij}^\parallel(\vec \eta \,) + \frac{\sinh^2 |\vec \eta \,|}{|\vec \eta \,|^2} P^\perp_{ij} (\vec \eta \,)\; ,
\ee
where $P^\parallel$ and $P^\perp$ are the parallel and perpendicular projectors in 
$\vec \eta$-space.
We can compute at once all one-loop contributions to the mass and to non-derivative interactions of the Goldstone fields by computing the one-loop Coleman-Weinberg potential \cite{CW}. Following again standard functional methods \cite{Weinberg},  we get
\be
\Delta \Gamma_{\rm CW} = \frac{i}{2} \int d^4 x \frac{d^4 k}{(2\pi)^4} {\rm tr} \log(k^2 f_{ij} (\vec \eta \,)) \, 
\ee
where the trace is a simple finite-dimensional (3$\times$3) matrix trace. We can split the matrix inside the trace as
\be
\log(k^2 f_{ij} (\vec \eta \,)) = \log(k^2) \delta_{ij} +  \log(f_{ij} (\vec \eta \,)) \; .
\ee
The first term is field independent, and we can discard it. As to the second term, we can evaluate its trace in a basis in which it is diagonal, such as a basis in which $\vec \eta \propto (1,0,0)$. We thus get
\be
\Delta \Gamma_{\rm CW} = i \Big(\int  \frac{d^4 k}{(2\pi)^4} \Big)  \int d^4 x \, \log \frac{\sinh^2 |\vec \eta \,|}{|\vec \eta \,|^2} = - \Delta S \; .
\ee
Regardless of the UV regulator used, this cancels exactly all effects of $\Delta S$ at this order, thus recovering agreement with the boost Goldstone theorem.

In conclusion, when using UV regulators other than dim-reg, the correction $\Delta S$ coming from the path-integral measure should be kept, and used consistently in perturbation theory.
In practice, for our purposes in this paper, this ends up not mattering. This is because we computed the one-loop expectation value of the stress-energy on the framid's ground state. Since $\Delta S$ is formally already of one-loop order, its contributions to such an expectation value should be considered only at tree level. That is,
\be
\langle T^{\mu\nu} \rangle_{\rm 1-loop} = \langle T^{\mu\nu} \rangle^0_{\rm 1-loop}
+ \langle \Delta T^{\mu\nu} \rangle_{\rm tree} \; ,
\ee
where the l.h.s.~stands for all one-loop contributions in the full theory (with action $S+\Delta S$), the first term on the r.h.s.~stands for the one-loop contributions in the theory without $\Delta S$, and the second term on the r.h.s.~stands for the correction to the stress-energy tensor operator coming from $\Delta S$, evaluated  on the ground state at tree level only. But at tree-level the ground state simply corresponds to $\vec \eta = 0$, so $\Delta S$ vanishes there, and so does its contribution to our expectation value.

%%%%%%%%%%%%%%%%%%%%%%%%%%%%%%%%%%%%%%%%%%%
%%%%%%%%%%%%%%%%%%%%%%%%%%%%%%%%%%%%%%%%%%%

%%%%%%%%%%%%%%%%%%%%%%%%%%%%%%%%%%%%%%%%%%%
%%%%%%%%%%%%%%%%%%%%%%%%%%%%%%%%%%%%%%%%%%%
% BIBLIOGRAPHY

\bibliography{library}{}
\bibliographystyle{utphys}

\end{fmffile}

\end{document}